\documentclass[12pt]{iopart}
\usepackage[dvips]{graphicx}
\usepackage{amssymb}
\usepackage{amsfonts}

\begin{document}

\title{Fluid and solid phases of the Gaussian core model}

\author{A Lang\dag\ddag, C N Likos\ddag,\footnote[3]{Corresponding author.
E-mail address: likos@thphy.uni-duesseldorf.de}
M Watzlawek{\ddag,}\footnote[4]{Present address:
Bayer AG, Central Research Division, D-51368 Leverkusen, Germany} and
H L\"owen\ddag}
\address{
\dag Institut f\"ur Theoretische Physik, Technische Universit\"at Wien,\\
Wiedner Hauptstra{\ss}e 8-10, A-1040 Wien, Austria\\
\ddag Institut f\"ur Theoretische Physik II, Heinrich-Heine-Universit\"at
D\"usseldorf,\\ 
Universit\"atsstra{\ss}e 1, D-40225 D\"usseldorf, Germany\\
}

\begin{abstract}
{We study the structural and thermodynamic properties of a model of
point particles interacting by means of 
a Gaussian pair potential first introduced by Stillinger
[Stillinger F H 1976 {\it J. Chem. Phys.} {\bf 65} 3968].
By employing integral equation theories for the fluid state and
comparing with Monte Carlo simulation results, we establish the limits
of applicability of various common closures and examine the dependence
of the correlation functions of the liquid on the density and
temperature. We employ a simple, mean-field theory for the high
density domain of the liquid and demonstrate that at infinite density
the mean-field theory is exact and that the system reduces to an
`infinite density ideal gas', where all correlations vanish and where
the hypernetted chain (HNC) closure becomes exact. By employing an
Einstein model for the solid phases, we subsequently calculate
quantitatively the phase diagram of the model and find that the
system possesses two solid phases, face centered cubic and 
body centered cubic, and also displays
reentrant melting into a liquid at high densities. Moreover, the
system remains fluid at all densities when the temperature exceeds
1\% of the strength of the interactions.}
\end{abstract}

\centerline
{Published: {\it J.\ Phys.: Condens.\ Matter} {\bf 12}, 5087 (2000)}


\section{Introduction\label{introduction}}

The structural and phase behaviour of systems 
whose constituent particles interact by means of pair
potentials diverging at zero separation between them,
is a problem that has been extensively studied in the last three decades,
by a variety of theoretical, experimental and computational 
methods. It has by now been
established that the diverging repulsions at close separations between
the particles are the dominant factor causing crystallisation of
the system; indeed, for the purposes of understanding the freezing
mechanism, the simple, hard sphere model is in most cases sufficient. 
The crystal structure into which a system freezes depends on the
steepness of the repulsion, with hard repulsions favouring a 
face centered cubic (fcc) lattice
and soft ones a body centered cubic
(bcc) lattice \cite{mcconnell:etal:prl:93}. 
On the other hand, interparticle
attractions are responsible for bringing
about a liquid-gas coexistence, whose stability with respect to
the freezing transition depends crucially on their relative range
with respect to that of their repulsive counterparts \cite{ilett:etal:pre:95}. 

Though repulsions are {\it necessary} to bring about a solidification
transition, they are by no means {\it sufficient}. This was demonstrated
recently by Watzlawek \etal \cite{watzlawek:etal:prl:99,martin:phd:00} 
who studied
a system interacting by means of an ultrasoft, logarithmically diverging
potential which has been shown to model accurately the effective 
interaction between star polymers in a good solvent
\cite{likos:etal:prl:98,jusufi:macromolecules:99}.
In particular, the pair potential employed in 
Refs.\ \cite{watzlawek:etal:prl:99,martin:phd:00}, reads as
\begin{eqnarray}
\nonumber
\beta v(r) & = & {{5}\over{18}}f^{3/2}
                 \left[-\ln\left({{r}\over{\sigma}}\right) + 
                 {{1}\over{1 + \sqrt{f}/2}}\right]\qquad\qquad\qquad
                 {\rm for}\;\; r \leq \sigma
\\
           & = & {{5}\over{18}}f^{3/2}{{1}\over{1 + \sqrt{f}/2}}
                 \left({{\sigma}\over{r}}\right)
                 \exp\left[-{{\sqrt{f}(r-\sigma)}\over{2\sigma}}\right]
                 \qquad{\rm for} \;\; r \geq \sigma,
\label{interaction.star}
\end{eqnarray}
where $\beta \equiv (k_BT)^{-1}$ is the inverse temperature ($k_B$ is
Boltzmann's constant), $f$ is the functionality of the stars and 
$\sigma$ is the typical extension of the stars,
the so-called corona diameter.
It was found that the strength of the repulsion, controlled by the
parameter $f$ in eq.\ (\ref{interaction.star}), is crucial in determining
whether the system crystallises. For values $f \leq 34$, the system
remains fluid at all densities and, even for values $f > 34$, 
reentrant melting into a high-density liquid has been observed for 
a particular range of functionalities 
\cite{watzlawek:etal:prl:99,martin:phd:00}.

We are, therefore, faced with a question which is specular to the old
question of how much attraction is needed in order to make a 
liquid \cite{frenkel:physicaa:99}, namely: {\it How much repulsion
is needed in order to make a solid?} In this respect, it is 
interesting to consider the extreme family of potentials that
are {\it bounded}, i.e., they remain finite for the whole range
of interparticle separations, even at full overlap between the 
particles. In the context of microscopic interactions in atomic
systems, such bounded potentials are evidently unphysical: full overlaps
between atoms, molecules or even compact macromolecular entities 
are forbidden by the Born repulsions between the electrons. Yet, they
are perfectly realistic in the context of {\it effective interactions}
between {\it fractal objects} such as polymer chains, as will be demonstrated
in section \ref{gaussian}.

In comparison with systems interacting by
means of unbounded potentials, very little is known about bounded
interactions. A model system that has been recently studied is the
`penetrable spheres model' (PSM), in which the interaction is a positive,
finite constant for separations smaller than a diameter $\sigma$ and
vanishing otherwise \cite{likos:pensph:98}. The model was studied
by means of 
cell-model calculations and
computer simulations \cite{likos:pensph:98},
liquid-state integral equation theories
\cite{fernaud:jcp:00} and density-functional
theory \cite{schmidt:cecam:99} and it was found that no reentrant
melting takes place with increasing density
because a clustering mechanism stabilises 
the solid at all temperatures. The PSM is, however, rather peculiar
in this respect and both the constant value of the potential inside
$\sigma$ and its sharp fall to zero outside $\sigma$ render it 
into a rather unrealistic model for effective interactions.

A much more realistic model of a bounded potential is the Gaussian
core model (GCM), introduced in the mid-seventies by Stillinger
\cite{stillinger:76}. In the GCM, the interaction potential between
the particles reads as
\begin{equation}
v(r) = \varepsilon e^{-(r/\sigma)^2},
\label{gaussian.potential}
\end{equation}
where $\varepsilon > 0$ is an energy and $\sigma$ a length scale.
Stillinger's original work \cite{stillinger:76}
was complemented later by molecular dynamics
simulations \cite{stillinger:weber:78,stillinger:weber:80}, high-temperature 
expansions \cite{stillinger:jcp:79} and the discovery 
of exact duality relations in the crystalline state \cite{stillinger:prb:79}.
Based on these considerations, a semi-quantitative phase diagram
of the GCM was drawn \cite{stillinger:76,stillinger:stillinger:97},
which displayed reentrant melting and a maximum freezing
temperature $t_{\rm u}$. The latter is defined as the temperature above
which the system remains fluid at all densities. 
Yet, a detailed study of the structure of the GCM fluid
by means of the modern techniques of liquid-state theory and a
{\it quantitative} calculation of the phase diagram of the GCM are
still lacking. The purpose of this work is to fill this gap and 
to draw from the study of the GCM some general conclusions, applicable
to a large class of bounded potentials.

The rest of the paper is organised as follows: in 
section \ref{gaussian} we briefly
review previous work on this model and make contact with
effective interactions between polymer chains. In 
section \ref{iets} we present
results about the structure of the fluid, obtained by means of
liquid-state integral equation theories and Monte Carlo simulations.
In section \ref{ideal}
we discuss the high density limit of the fluid, by employing
a simple mean-field density functional. In 
section \ref{einstein} we present the
theory for the crystal phases and in section \ref{phdg}
we make use of the
results for the fluid and the solids in order to quantitatively draw
the phase diagram of the system. Finally, in section \ref{summary}
we summarise and conclude.

\section{Gaussian effective interactions\label{gaussian}}

A Gaussian pair potential was first proposed fifty years ago
by Flory and Krigbaum \cite{flory:krigbaum:50},
as the effective interaction
between the centres of mass of two polymer chains. This so-called
Flory-Krigbaum potential, $v_{\rm FK}(r)$, reads as
\begin{equation}
\beta v_{\rm FK}(R) = N^2 {{v^2_{\rm seg}}\over{v_{\rm solv}}}
\left({{3}\over{4\pi R^2_g}}\right)^{3/2}(1-2\chi)
\exp\left(-{{3}\over{4}}{{R^2}\over{R^2_g}}\right),
\label{vfk}
\end{equation}
where $v_{\rm seg}$ and $v_{\rm solv}$ denote the volumes of a monomer
segment and a solvent molecule, respectively, $N$ is the degree of
polymerisation, $R_g$ the radius of gyration of the chains and
$\chi$ is a parameter that controls the solvent quality
($0 < \chi < 1/2$ denotes a good solvent and $\chi > 1/2$ a poor one.) 

Formally, the effective interaction potential $v_{\rm eff}(r)$
between two polymer chains is defined as follows:
the
centres of mass of the two chains are held fixed at separation $r$, and
the canonical trace is taken over all the monomer degrees of freedom.
In this way, a restricted partition function $Q(r)$ is obtained,
where the restriction denotes the constraint of holding the centres
of mass fixed. The effective interaction is then simply the free
energy $-k_BT\ln Q(r)$. As can be seen in Fig.\ \ref{twochains.plot},
the centres of mass of two polymer chains can coincide without any
violation of the excluded volume condition between the monomers (beads).
Hence, the effective interaction between polymer centres of mass 
does not have to diverge at $r = 0$ and a bounded potential
is fully realistic.
\begin{figure}[hbt]
\begin{center}
\end{center}
\caption{A snapshot from a simulation involving two self-avoiding
         polymers. In this
         configuration, the centres of mass of the two chains
         (denoted by the big
         sphere) coincide, without violation of the excluded-volume
         conditions.
         (Courtesy of Arben Jusufi.)}
   \label{twochains.plot}
\end{figure}

The effective interaction between linear polymers
has been extensively studied by computer simulations, both
on-lattice \cite{olaj:pelinka:76,olaj:lantschbauer:pelinka:80}
and off-lattice 
\cite{olaj:lantschbauer:77,grosberg:82,schaefer:baumgaertner:86,
krueger:etal:89,dautenhahn:hall:94}
and for a varying solvent quality. In addition, Kr{\"u}ger \etal
\cite{krueger:etal:89}
performed a theoretical calculation of
the effective interaction for athermal solvents, using renormalisation-group
techniques. Regardless of the detailed manner in which the monomer-monomer
excluded volume interactions have been modeled, all these studies converge
to the result that 
the effective potential $v_{\rm eff}(r)$ is extremely well
approximated by a {\it universal} Gaussian function\footnote{The
universality holds 
when the two chains
have the same degree of polymerisation and the solvent is athermal.
If the temperature becomes relevant or the two chains have
different numbers of monomers, corrections to the Gaussian form
appear.} given by eq.\ (\ref{gaussian.potential}),
with $\varepsilon$ being an energy scale of the order of the thermal
energy $k_BT$ and $\sigma$ a length scale of the order of the
spatial extent of the chains (e.g., the radius of gyration). 
This type of effective
Gaussian interaction has been recently employed by 
Louis \etal \cite{ard:finken:hansen:99} 
in their study of the structure of colloid-polymer
mixtures. 

The Gaussian effective potential between polymer chains in purely
entropic in nature, causing the energy $\varepsilon$ to scale with 
temperature. Therefore, the latter completely drops out of the problem
as a relevant thermodynamic parameter. However, since the system is
of significant theoretical interest, it is pertinent to consider
$\varepsilon$ and $k_BT$ in more generality, as two independent 
energy scales and study the properties of the system as functions
of density and temperature. To this goal, we define the dimensionless
temperature $t$ as
\begin{equation}
t \equiv {{k_BT}\over{\varepsilon}} = \left(\beta\varepsilon\right)^{-1}.
\label{dimless.t}
\end{equation}
The density $\rho$ of a system of $N$ particles 
(or polymer chains in the above discussion) enclosed in the macroscopic
volume $V$ 
is the ratio $N/V$.
As dimensionless measures of the density, we are going to use 
to use in this paper both the parameter $\bar\rho \equiv \rho\sigma^3$
and the `packing fraction' $\eta$, defined as
\begin{equation}
\eta \equiv {{\pi}\over{6}}\rho\sigma^3.
\label{eta}
\end{equation}

A number of exact properties for the GCM were worked out by Stillinger
\etal \cite{stillinger:76,stillinger:prb:79,stillinger:stillinger:97},
yielding important information on the topology of
the phase diagram. The relevant properties are listed below.

{\it 1. Hard sphere limit.} At low temperatures and densities, 
the GCM reduces to a hard sphere (HS) system with a HS diameter 
that diverges as the temperature approaches zero \cite{stillinger:76}.
Hence, the GCM displays there the usual HS freezing
transition from a fluid into a fcc lattice and the coexistence 
densities converge to zero at vanishing temperature. Using the
known results for the HS 
freezing \cite{alder:hoover:young:68,hoover:ree:68,haymet:oxtoby:86}
the shape of the freezing and
melting lines $t_{\rm f}(\bar \rho)$ and
$t_{\rm m}(\bar \rho)$ as $t \rightarrow 0$ is given by the equations
\begin{eqnarray}
t_{\rm f}(\bar\rho) & = & {{1}\over{\ln 2}}
\exp\left(-{{0.962}\over{\bar\rho^{\;2/3}}}\right),
\label{freezing.line}
\\
t_{\rm m}(\bar\rho) & = & {{1}\over{\ln 2}}
\exp\left(-{{1.027}\over{\bar\rho^{\;2/3}}}\right).
\label{melting.line}
\end{eqnarray}

{\it 2. Duality relations.} Due to the property of the Gaussian potential
to remain form-invariant under Fourier transformation, it has been shown
that the internal energy of a certain crystal at low densities
(lattice sum) is related to that of the reciprocal crystal at 
high densities \cite{stillinger:prb:79}. In this way it was established
that the fcc and bcc lattices are degenerate at $t = 0$ at the 
`magic density' $\bar\rho_{*} = \pi^{-3/2} \cong 0.1796$, with the
fcc ``winning'' for $\bar\rho < \bar\rho_{*}$ and the bcc for 
$\bar\rho > \bar\rho_{*}$.

{\it 3. Reentrant melting.} At small but finite temperatures and at 
high densities the bcc lattice remelts into a fluid and the high density
freezing and melting lines are given by the 
equations \cite{stillinger:prb:79}
\begin{eqnarray}
t_{\rm {f}}(\bar\rho) & \propto \exp\left(-K_{\rm {f}}\bar\rho^{\;2/3}\right),
\label{freeze.highrho}
\\
t_{\rm {m}}(\bar\rho) & \propto \exp\left(-K_{\rm {m}}\bar\rho^{\;2/3}\right),
\label{melt.highrho}
\end{eqnarray}
with the appropriate constants $K_{\rm {f}}$ and $K_{\rm {m}}$. Note
the duality between eqs.\ (\ref{freezing.line}), (\ref{melting.line}),
valid for $\bar\rho \to 0$, and (\ref{freeze.highrho}),
(\ref{melt.highrho}), valid for
$\bar\rho \to \infty$. Apart from proportionality constants, one pair
can be obtained from the other by the formal substitution
$\bar\rho \rightarrow 1/\bar\rho$. Moreover, eqs.\ (\ref{freeze.highrho})
and (\ref{melt.highrho})
show that the slope of the high-density freezing and melting lines are
{\it negative}. Contrary to the usual case, the liquid coexisting with
the solid has a {\it higher} density than the latter.  

The above general considerations were combined with molecular dynamics
simulations \cite{stillinger:weber:78,stillinger:weber:80} 
and an approximate phase
diagram of the model was drawn \cite{stillinger:stillinger:97}. 
It was found that 
below an `upper freezing temperature' $t_{\rm u} \approx 0.008$,
the GCM displays the transitions fluid $\to$ fcc, fcc $\to$ bcc
and bcc $\to$ fluid with increasing density, whereas above $t_{\rm u}$
a {\it single} fluid phase exists and no freezing takes place,
at any density.
However, the liquid structure of the GCM has not been studied to-date
and the phase diagram was drawn 
only semi-quantitatively. In the following
sections we perform a quantitative analysis on all these aspects,
using standard tools of statistical mechanics.  

\section{The fluid: integral equation theories and Monte Carlo
simulations\label{iets}}

The theoretical determination of the pair structure of a uniform
fluid amounts to the calculation of the radial distribution 
function $g(r)$ and the direct correlation function (dcf) $c(r)$
\cite{hansen:mcdonald}. The correlation function $h(r)$ is 
simply $g(r) - 1$ and then $h(r)$ and $c(r)$ are 
connected to each other through the {\it Ornstein-Zernike relation}
which has the form \cite{hansen:mcdonald}
\begin{equation}
h(r) = c(r) + \rho \int d{\bf r'} c(|{\bf r} - {\bf r'}|) h(r').
\label{oz}
\end{equation}
With $\tilde h(Q)$ denoting the Fourier transform of $h(r)$, the
structure factor $S(Q)$ is defined as
\begin{equation}
S(Q) = 1 + \rho \tilde h(Q).
\label{sofq.define}
\end{equation}

The Ornstein-Zernike relation is exact. As it connects two unknown functions,
one more relation or {\it closure}
is needed in order to determine $g(r)$ and $c(r)$. Closures are approximate
relations which arise from exact diagrammatic expansions of $g(r)$
in terms of $c(r)$ but with certain classes of diagrams ignored.
The exact relation between $g(r)$ and $c(r)$ reads as \cite{hansen:mcdonald}
\begin{equation}
g(r) = \exp\left[-\beta v(r) + h(r) - c(r) + B(r)\right],
\label{gandc.exact}
\end{equation}
where $v(r)$ is the pair potential and
$B(r)$ is the so-called {\it bridge function}, consisting of the sum
of all elementary diagrams that are not nodal.

All known closures can be
thought of as approximate relations for the form of $B(r)$.
Common closures are the Percus-Yevick (PY) and Hypernetted Chain (HNC)
approximations \cite{hansen:mcdonald}. In the PY closure, the approximation
for $B(r)$ reads as
\begin{equation}
B_{\rm PY}(r) =  c(r) - h(r) + \ln[g(r)-c(r)],
\label{bridge.py}
\end{equation}
whereas in the HNC the approximation is made that the bridge function vanishes:
\begin{equation}
B_{\rm HNC}(r) = 0. 
\label{bridge.hnc}
\end{equation} 

The PY closure is successful only for short-ranged, hard interactions
and hence it will not be considered here. In addition to the HNC, we
have considered two more closures of increasing degree of sophistication,
the Rogers-Young (RY) closure \cite{rogers:young}
and the zero-separation (ZSEP) 
closure \cite{verlet:zsep:80,lee:shing:89,lee:92,lee:95,lee:ghonasgi:lomba:96}.
The former (RY) closure reads as
\begin{equation}
g(r) = \exp\left[-\beta v(r)\right]
\left[1 + {{\exp\left[\gamma(r)f(r)\right] - 1}\over{f(r)}}\right],
\label{ry}
\end{equation}
where
\begin{equation}
\gamma(r) = h(r) - c(r)
\label{gammaofr}
\end{equation}
and the `mixing function' $f(r)$ is chosen to have the form
\begin{equation}
f(r) = 1 - \exp\left(-\alpha r\right).
\label{mixing.function}
\end{equation}
The parameter $\alpha$ is determined so that thermodynamic consistency
between the virial and compressibility pressures is 
achieved \cite{hansen:mcdonald}. In the limit $\alpha = 0$ one
recovers the PY and in the opposite limit, $\alpha \to \infty$,
the HNC closure.

The ZSEP closure includes three
parameters and is a direct approximation for the bridge function
having the form
\begin{equation}
B_{\rm ZSEP}(r) = -{{\zeta \gamma^2(r)}\over{2}}
                   \left[1 - {{\varphi \gamma(r)}\over{1 + \alpha \gamma(r)}}
                   \right].
\label{bzsep}
\end{equation}
The ZSEP closure has been applied recently by 
Fernaud \etal \cite{fernaud:jcp:00} to the PSM model mentioned above,
yielding excellent results for the correlation functions as compared
to the simulation results.
The three
parameters $\zeta$, $\varphi$ and $\alpha$ are determined in such a way
that virial-compressibility, Gibbs-Duhem and zero separation consistencies
are all fulfilled; for details on the above we refer the reader to
Ref.\ \cite{fernaud:jcp:00}. Evidently, for $\zeta = 0$ the ZSEP
reduces to the HNC. 

\begin{figure}[hbt]
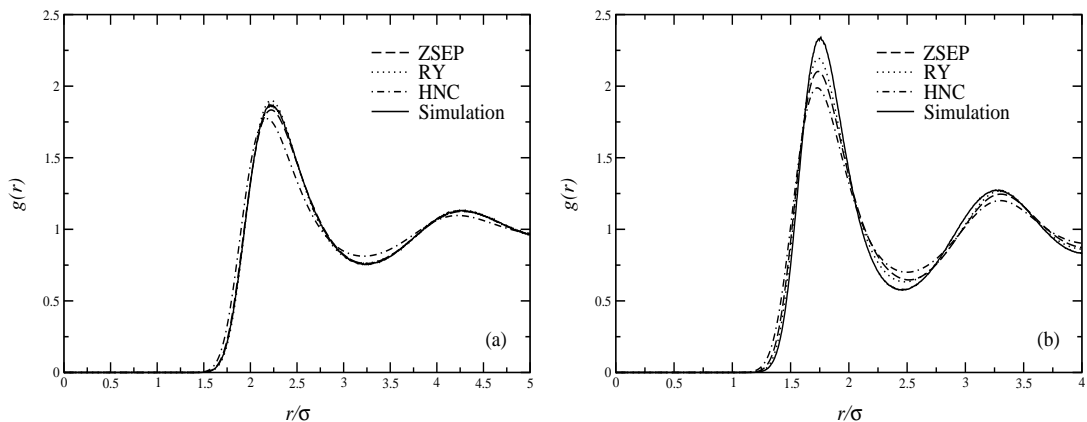

\begin{center}
\begin{minipage}[t]{7.2cm}
   \includegraphics[width=7.0cm,height=5.5cm,clip]
   {e0.05.eps}
   \end{minipage}
   \begin{minipage}[t]{7.2cm}
   \includegraphics[width=7.0cm,height=5.5cm,clip]
   {e0.12.eps}
   \end{minipage}
   \end{center}
\caption{Comparison for the radial distribution function $g(r)$
between the simulation results and those obtained
from the various closures, at $t = 0.01$ and
at small values of the packing fraction.
(a) $\eta = 0.05$; (b) $\eta = 0.12$.}
   \label{small.gr.plot}
\end{figure}

\begin{figure}[hbt]
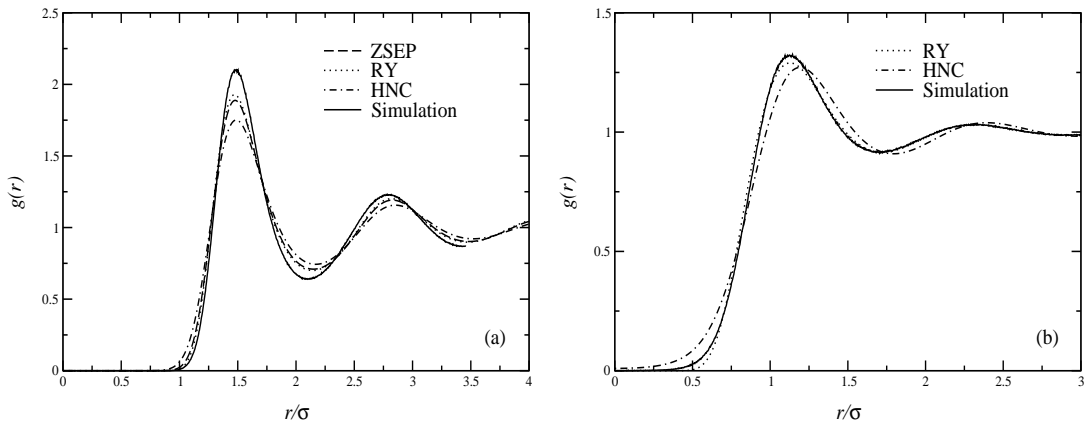

\begin{center}
\begin{minipage}[t]{7.2cm}
   \includegraphics[width=7.0cm,height=5.5cm,clip]
   {e0.20.eps}
   \end{minipage}
   \begin{minipage}[t]{7.2cm}
   \includegraphics[width=7.0cm,height=5.5cm,clip]
   {e0.50.eps}
   \end{minipage}
   \end{center}
\caption{Same as Fig.\ \ref{small.gr.plot} but for intermediate packings.
(a) $\eta = 0.20$; (b) $\eta = 0.50$. Note the different scale of the 
    vertical axes.}
   \label{med.gr.plot}
\end{figure}

\begin{figure}[hbt]
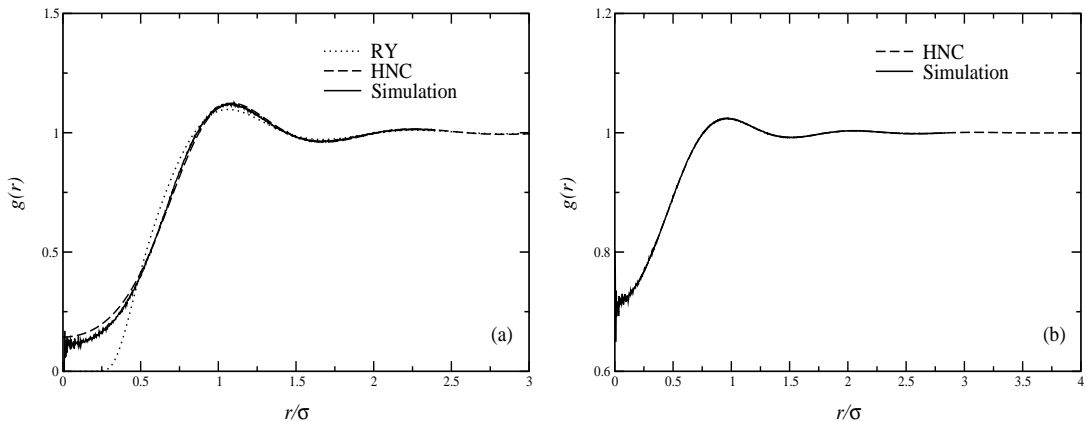

\begin{center}
\begin{minipage}[t]{7.2cm}
   \includegraphics[width=7.0cm,height=5.5cm,clip]
   {e1.00.eps}
   \end{minipage}
   \begin{minipage}[t]{7.2cm}
   \includegraphics[width=7.0cm,height=5.5cm,clip]
   {e6.00.eps}
   \end{minipage}
   \end{center}
\caption{Same as Fig.\ \ref{small.gr.plot} but for high packings. 
(a) $\eta = 1.00$; (b) $\eta = 6.00$, where the simulation result
is indistinguishable from the HNC.}
   \label{big.gr.plot}
\end{figure}

In addition, in order to test the reliability of the various closures,
we have performed standard $NVT$ Monte Carlo (MC) simulations for different
densities and temperatures. In what follows,
we focus our attention to temperature
$t = 0.01$ on two grounds: on the one hand, according to the
approximate phase diagram \cite{stillinger:stillinger:97}, at $t = 0.01$
the system remains fluid at all densities and therefore we can study
the development of the correlation functions with density for an
unlimited range of the latter, without entering a region where the
fluid is metastable.
On the other hand,
this temperature is low
enough, so that significant structure is expected for the correlation
functions of the fluid and hence the integral equation theories
can be put in a strong test. 
We present the obtained results in 
Figs.\ \ref{small.gr.plot}, \ref{med.gr.plot} and 
\ref{big.gr.plot} and we discuss them below.

We begin with the RY closure. When the packing is not large enough,
typically $\eta \lesssim 0.50$ at this temperature, the RY closure
gives results which are in very good agreement with simulations,
as can be seen in Figs.\ \ref{small.gr.plot} and \ref{med.gr.plot}. 
However, above $\eta \approx 0.50$, the $g(r)$ from simulations
starts {\it penetrating} towards $r = 0$, physically meaning that
the probability of finding two particles `sitting on top of each
other' is finite and there is no `correlation hole' in $g(r)$.
This is to be expected for a system with a bounded interaction.
However,
as can be seen from Fig.\ \ref{big.gr.plot} (a), the RY closure
fails to capture precisely this penetration phenomenon, yielding
$g(r)$'s that are {\it too low} at small separation and
making the erroneous prediction 
that a correlation hole exists. 

The reason for this behaviour can be
traced back in the construction of the RY closure, eqs.\ 
(\ref{ry}) and (\ref{mixing.function}), where it can be seen that
the RY closure always looks like the PY closure at small separations.
The latter is however inaccurate for a long-range interaction lacking
a hard core. We attempted to modify the RY closure by employing
mixing functions yielding a HNC-like small-$r$ behaviour and a 
PY-like large-$r$ behaviour. However, this did not yield self-consistency
parameters for the whole range of densities. Despite of its inability
to correctly describe the high-density penetration of $g(r)$, the
standard RY closure yields nevertheless a self-consistency parameter
$\alpha$ for all densities considered here. In addition, this parameter
grows with density, thus pointing to a tendency of the RY closure
to reduce to the HNC at high packings.
      
\begin{figure}[hbt]
\begin{center}
\includegraphics[width=9.0cm,height=7.5cm,clip]
{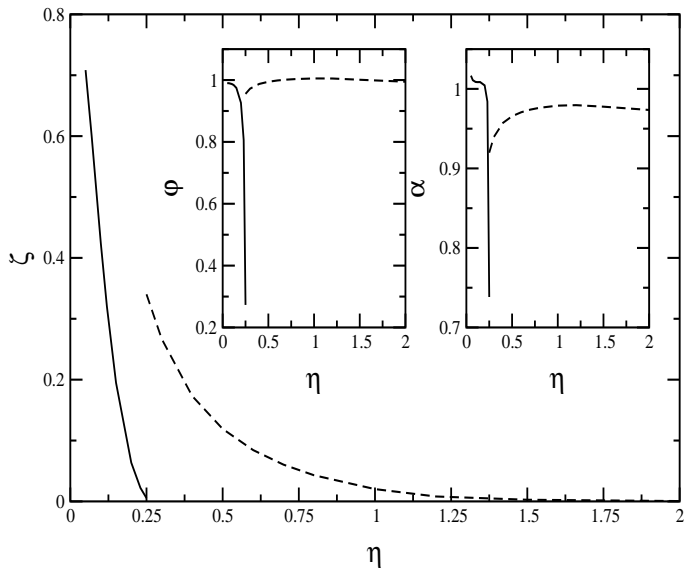}
\end{center}
\caption{The self-consistency parameters of the ZSEP closure applied
to the Gaussian core model at reduced temperature $t = 0.01$ as functions
of the packing fraction $\eta$. The solid lines denote the physical
and the dashed lines the unphysical branch (see the text.)}
   \label{para.plot}
\end{figure}

In view of the failure of the RY-closure, we are led to consider
the ZSEP closure which has {\it precisely} the property that 
in its implementation the value
of $g(r)$ at zero separation plays an important role and is determined
self-consistently. In fact, the resounding success of the ZSEP in
describing the $g(r)$ of the PSM model (also a bounded interaction)
has been mainly attributed to this property \cite{fernaud:jcp:00}. 
As can be seen from Figs.\ \ref{small.gr.plot}
and \ref{med.gr.plot}(a), the ZSEP performs only slightly worse than
the RY closure up to a packing fraction $\eta \approx 0.25$. The 
self-consistency parameters $\zeta$, $\varphi$ and $\alpha$ of the ZSEP
are displayed in Fig.\ \ref{para.plot}. The parameter $\zeta$ which
appears as an overall factor in the parametrisation of the bridge
function [see eq.\ (\ref{bzsep})] decreases with density and close
to $\eta = 0.25$ it is small enough and the ZSEP practically reduces to
the HNC closure.

However, at packing fractions $\eta > 0.25$, a {\it second branch} of
solutions appears, which is denoted by the dashed lines in 
Fig.\ \ref{para.plot}. This branch is
disconnected from the first and hence it causes the 
$g(r)$ to behave discontinuously at this packing fraction,
a result which is clearly unphysical. This second branch produces
$g(r)$'s that show {\it too much penetration} and too little 
structure, when compared with the simulation results. The reason for
this unphysical behaviour can be traced to the fact that this second
branch corresponds to bridge functions which are {\it positive}
at small separations $r$. Indeed, from eq.\ (\ref{bzsep}), and 
taking into account that $\gamma(r) \gg 1$ at small separations,
we can see that $B_{\rm ZSEP}(r)$ has the sign of the product
$\zeta(\varphi - \alpha)/\alpha$. For the first, physical branch 
(solid lines in Fig.\ \ref{para.plot}) this combination is
{\it negative} because $\varphi < \alpha$. For the second,
unphysical branch, this combination is positive because $\varphi > \alpha$
there. A positive bridge function acts then as an additional
`attractive potential' in eq.\ (\ref{gandc.exact}) and causes the
overpenetration in $g(r)$ mentioned above.

The appearance of a branch of solutions of the ZSEP for which the
bridge function is positive is a sign of internal inconsistency
of the closure and in this sense the ZSEP signals its own limits 
of applicability. Indeed, the bridge function of any system has
been shown to be a quasi-universal function, which can always
be mapped onto the bridge function of a suitably defined
hard sphere system having a hard sphere diameter that depends on
the characteristics of the interaction and the thermodynamic 
point under consideration \cite{yasha:nwa:mhnc}. The bridge function
of the HS system is, however, practically exactly known and 
it is essentially a negative function for all $r$.
Hence, a positive bridge function
is physically unacceptable and the second branch of solutions 
of the ZSEP has to be discarded. The results from this second branch
come again into very good agreement with simulation at packing
fractions above $\eta \approx 1.00$ because, as seen in 
Fig.\ \ref{para.plot}, the parameter $\zeta$ is already very small
there and the bridge function has a negligible contribution to 
$g(r)$; even the unphysical branch reduces to the HNC at high 
densities. However, there is no way to bridge the physical 
solutions at packings $\eta \leq 0.25$ with the HNC-like solutions
{\it within} the ZSEP closure, that is without having to compare
with independently produced simulation results.

\begin{figure}[hbt]
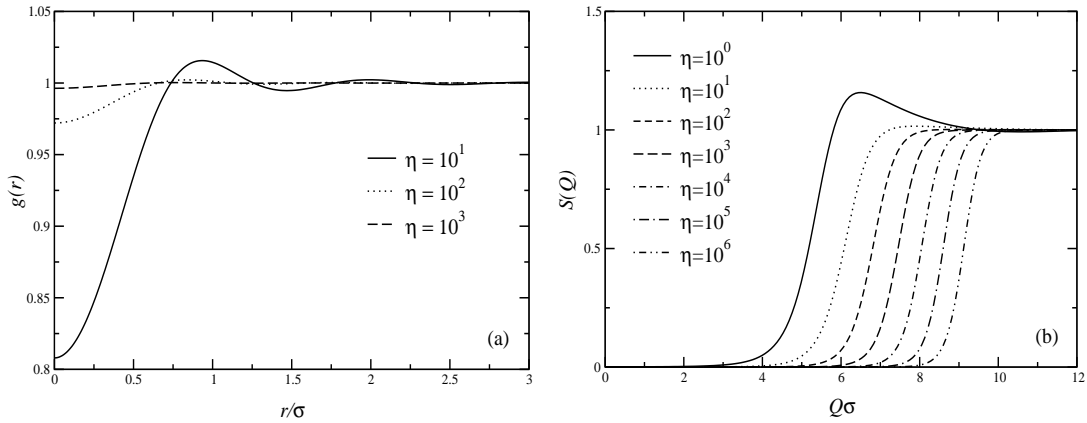

\begin{center}
\begin{minipage}[t]{7.2cm}
   \includegraphics[width=7.0cm,height=5.5cm,clip]
   {huge.gr.eps}
   \end{minipage}
   \begin{minipage}[t]{7.2cm}
   \includegraphics[width=7.0cm,height=5.5cm,clip]
   {scomp.eps}
   \end{minipage}
   \end{center}
\caption{(a) The radial distribution function $g(r)$ and (b) the 
structure factor $S(Q)$ as obtained by the HNC closure for
extremely high values of the packing fraction.}
   \label{huge.eta.plot}
\end{figure}

We comment next on the quality of the HNC. As can be seen from
Figs.\ \ref{small.gr.plot} - \ref{big.gr.plot}, the HNC underestimates
the structure at small to intermediate packings but yields otherwise
reasonable results. It does not suffer from any of the problems
of the more sophisticated closures and, in fact, it seems to be the
best at high densities. In order to further explore this property
(which is supported by the fact that the other two closures tend
to the HNC at this limit), we also solved this closure at 
{\it extremely high} packing fractions. Here, the highest packing
at which we simulated was $\eta = 6.00$, due to time constraints. 
With increasing $\eta$, a very large number of particles
would be required in the simulation box in order to obtain 
reliable results. In view of the fact that the HNC gives results
which coincide with those from Monte Carlo at $\eta = 6.00$
[see Fig.\ \ref{big.gr.plot}(b)],
we gain confidence at this closure and apply it 
for arbitrarily high densities, in order to obtain information 
on the structure of the very dense fluid. In Fig.\ \ref{huge.eta.plot}(a),
we show results for $g(r)$ where it can be seen that at very high
packings $g(r)$ tends to unity and hence $h(r) = g(r) - 1$ tends to zero.
However, this does not mean that
the structure factor $S(Q)$ tends to unity as well, as a naive guess
based on the definition $S(Q) = 1 + \rho\tilde h(Q)$ would imply.
The quantity $\tilde h(Q)$ tends to zero but at the same time the
prefactor $\rho$ diverges, thus giving rise to a nontrivial $S(Q)$.

Results for the corresponding structure factor $S(Q)$ are shown
In Fig.\ \ref{huge.eta.plot}(b).  
It can be seen that, for high densities, the peak of $S(Q)$ disappears
and the latter looks like a `smoothed step function' with values
ranging from zero to one. The value of $Q$ at which the crossover
occurs does not scale as a power-law of the density but
rather shifts to the right by almost
a constant every time the packing fraction is
increased by an order of magnitude. This hints to a very weak
dependence of this crossover value on the density; we return to this
point in section \ref{ideal}.

\begin{figure}[hbt]
\begin{center}
\includegraphics[width=9.0cm,height=7.5cm,clip]
{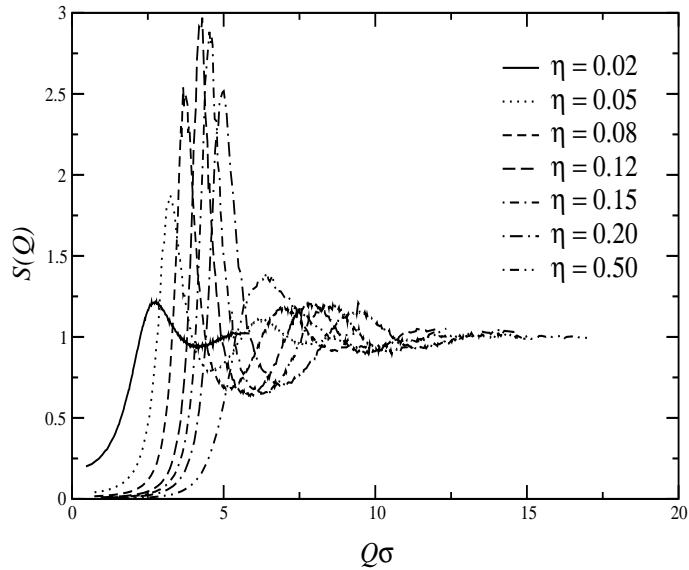}
\end{center}
\caption{The structure factor $S(Q)$ of the GCM at $t = 0.01$ for
         different packing fractions, as obtained by Monte Carlo 
         simulation.}
   \label{ssim.plot}
\end{figure}

The liquid-state correlation functions of the GCM
display an anomalous behaviour
in comparison with that of `normal' liquids, interacting by means of
hard, unbounded interactions. For the latter, the structure grows
monotonically with increasing density and eventually the systems
freeze. Here, the structure grows up to a packing fraction 
$\eta \cong 0.12$ at $t = 0.01$ and for higher densities it becomes
weaker again. To amply demonstrate this phenomenon we show in 
Fig.\ \ref{ssim.plot} the structure factors obtained from the
MC simulations for a wide range of densities, where it
can be seen that the height of the peak of $S(Q)$ attains its maximum
value at $\eta \cong 0.12$. The same phenomenon has been observed in
the liquid structure of star polymer solutions employing the
interaction potential given by eq.\ (\ref{interaction.star})
\cite{likos:etal:prl:98,martin:phd:00,watzlawek:etal:jpcm:98} and, in fact, 
in star polymer solutions the effect has also been
observed experimentally \cite{olaf:phd:95}.
This behaviour of $S(Q)$ is closely related to reentrant 
melting \cite{watzlawek:etal:prl:99}. Indeed, the height of the maximum
of $S(Q)$ is a diagnostic tool for the freezing transition. According
to the Hansen-Verlet criterion \cite{hansen:verlet:69,hansen:schiff:73},
a liquid crystallises when
$S(Q)$ at its highest peak has the quasi-universal value 2.85. At higher
values of the peak the system is solid and at lower fluid. The evolution
of $S(Q)$ with density shown in Fig.\ \ref{ssim.plot} in conjunction
with the Hansen-Verlet criterion implies that the system approaches
crystallization at about $\eta = 0.12$ and then remelts. The height
of the peak at $\eta = 0.12$, which is indeed slightly above the
Hansen-Verlet value, implies that at this temperature the system 
barely freezes and that $t = 0.01$ is very close to the upper freezing
temperature. We will confirm this prediction in section \ref{phdg},
where we will also use the structural results obtained here in order
to calculate the free energy of the fluid.

Finally, we have performed MC simulations and solved the HNC closure
also at much higher temperatures,
$t = 1.00$, corresponding to the physical domain for the effective 
interactions between polymer chains. There, we found that the HNC
accurately reproduces the simulation results at all densities and
that the liquid has very little structure, a result which can be
easily understood in view of the fact that the thermal energy,
which is equal to the interaction strength there,  
washes out the correlation effects caused by the interactions.

\section{The high-density limit in the fluid state\label{ideal}}

The results of the preceding section on the structure of the
fluid at high densities point out that the correlations are becoming
weaker as density grows and that the system approaches some kind of
`infinite density ideal gas' limit, where $g(r) = 1$. This limit
was assumed already by Stillinger \cite{stillinger:76}.
There, the internal energy of a high-density fluid was
approximated by that of a {\it random distribution} of
points interacting by means of the Gaussian potential,
i.e., the positions of the points were assumed
to be uncorrelated. The relation $g(r) = 1$ was used explicitly
there in deriving an estimate for the internal energy of the high-density
fluid and it was shown that in fact the complete absence of a
correlation hole in the fluid raises its internal energy with respect
to that of a solid (where a correlation hole is necessarily present)
by exactly $\varepsilon/2$ per particle. Yet, the existence of this
ideal-gas limit was not proven.
Here, we are going to present
a density functional mean-field theory which establishes this limit
and provides analytic expressions for the correlation functions of
the GCM fluid at high densities.

In the framework of density functional theory (DFT), one considers
in general a system with a spatially modulated one-particle
density $\rho({\bf r})$. The Helmholtz free energy of the system is then
a {\it unique functional} of the one-particle density \cite{evans:79},
$F = F[\rho({\bf r})]$, and
can be written as a sum of an entropic, ideal part $F_{\rm id}[\rho({\bf r})]$
which is exactly known and an excess part $F_{\rm ex}[\rho({\bf r})]$
which is in general unknown:
\begin{eqnarray}
\nonumber
F[\rho({\bf r})] & = & F_{\rm id}[\rho({\bf r})] + F_{\rm ex}[\rho({\bf r})]
\\
& = &
k_B T\int d{\bf r} \rho({\bf r}) \Bigl[\ln\bigl[\rho({\bf r}) \Lambda^3\bigr]
-1\Bigr] + F_{\rm ex}[\rho({\bf r})].
\label{ideal.excess}
\end{eqnarray}
In eq.\ (\ref{ideal.excess}) above, $\Lambda$ denotes the thermal
de Broglie wavelength of the particles.

We consider the limit $\rho\sigma^3 \gg 1$. The average interparticle
distance $a \equiv \rho^{-1/3}$ becomes vanishingly small in this 
limit and it holds $a \ll \sigma$, i.e., the potential is extremely
long-range. Every particle is simultaneously interacting with
an enormous number of neighbouring molecules and in the absence of
short-range excluded volume interactions the excess free energy
of the system can be approximated by a simple mean-field term, equal to
the internal energy of the system: 
\begin{equation}
F_{\rm ex}[\rho({\bf r})] \cong 
{{1}\over{2}} \int\int d{\bf r} d{\bf r'} v(|{\bf r} - {\bf r'}|)
\rho({\bf r}) \rho({\bf r'}),
\label{dft.mfa}
\end{equation}
with the approximation becoming more accurate with increasing density
and eventually exact at $\bar\rho \to \infty$. The direct correlation
function $c(|{\bf r} - {\bf r'}|;\rho)$ in a fluid of density $\rho$
is given by the second functional derivative of 
$F_{\rm ex}[\rho({\bf r})]$ with
respect to the density \cite{evans:79}, namely
\begin{equation}
c(|{\bf r} - {\bf r'}|;\rho) = 
-\lim_{\rho({\bf r}) \to \rho}
{{\delta^{2} \beta F_{\rm ex}[\rho({\bf r})]}\over
 {\delta \rho({\bf r}) \delta \rho({\bf r'})}}.
\label{dcf.dft}
\end{equation}
Combining eqs.\ (\ref{dft.mfa}) and (\ref{dcf.dft}) we find that 
the dcf of the system at high densities becomes independent of
$\rho$ and is simply proportional to the interaction:
\begin{equation}
c(r) = -\beta v(r).
\label{msa}
\end{equation}

Eq.\ (\ref{msa}) above has a strong similarity with the mean-spherical
approximation (MSA) \cite{lebowitz:percus:66,hansen:mcdonald}, introduced
as a perturbation theory to study systems interacting by potentials  
that can be separated into a hard sphere
interaction (diameter $\sigma_{\rm HS}$)
$v_0(r)$ and a `soft tail' $\phi(r)$. In the MSA, the radial distribution
function of the system $g(r)$ vanishes for $r < \sigma_{\rm HS}$ and
the direct correlation function $c(r)$ at $r > \sigma_{\rm HS}$ is
given by $c(r) = -\beta \phi(r)$. The main difference from the 
MSA here, is that there is {\it no reference potential} $v_0(r)$
because there are no hard cores in the system. Hence, eq.\ (\ref{msa})
holds with the {\it total} interaction on the right hand side
and for the {\it whole range} of separations $r$. Moreover,
unlike the MSA which is essentially a high-temperature approximation,
eq.\ (\ref{msa}) holds for the whole temperature range, provided 
that the density is high enough ($\rho\sigma^3 \gg 1$). Of course,
the mean-field approximation becomes also valid at high temperatures
($t \gg 1$) irrespective of the density, because there the thermal
energy completely dominates over the {\it bounded} interaction.
This happens in contradistinction with diverging interactions,
where short-range correlation effects always survive, at all
temperatures.

\begin{figure}[hbt]
\begin{center}
\includegraphics[width=9.0cm,height=7.5cm]
{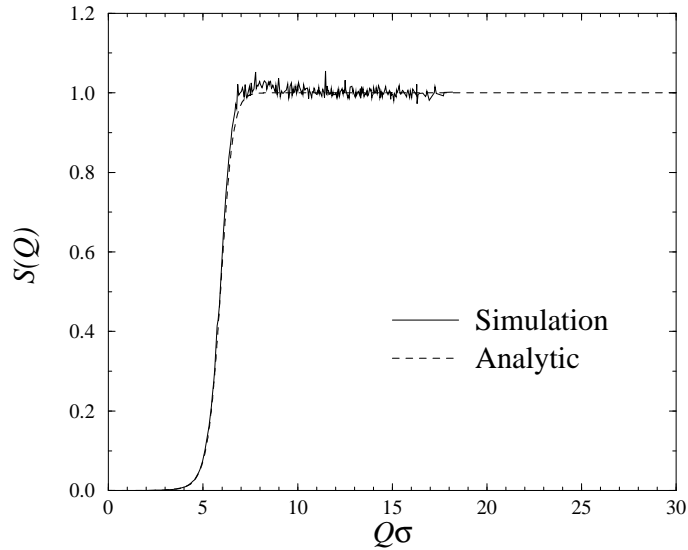}
\end{center}
\caption{The structure factor $S(Q)$ of the GCM at $t = 0.01$ and
         $\eta = 6.00$ as obtained from simulation (solid line) and
         as given by the analytical expression, eq.\ (\ref{sofq.analytic})
         (dashed line).}
   \label{sofq.ideal.plot}
\end{figure}

The Fourier transform $\tilde c(Q)$ of the dcf is obtained easily
from eqs.\ (\ref{msa}) and (\ref{gaussian.potential}), 
and for the GCM it has the form:
\begin{equation}
\tilde c(Q) = -\pi^{3/2}\beta\varepsilon\sigma^3
              \exp\left[-\left( Q\sigma/2 \right)^2\right].
\label{cofq}
\end{equation}
Using the Ornstein-Zernike equation and the ensuing relation 
$S(Q) = [1 - \rho \tilde c(Q)]^{-1}$ we obtain for the structure factor
of the GCM the expression:
\begin{equation}
S(Q) = {{1}\over{1 + \pi^{3/2}\beta\varepsilon\rho\sigma^3
                 \exp\left[-\left( Q\sigma/2 \right)^2\right]}}.
\label{sofq.analytic}
\end{equation}
This analytic expression is compared with the MC result at
$\beta\varepsilon = 100$ ($t = 0.01$) and $\eta = 6.00$ in 
Fig.\ \ref{sofq.ideal.plot}. The excellent agreement between the 
two demonstrates the validity of the simple mean-field theory at
high densities. 
Eq.\ (\ref{sofq.analytic}) shows that at high densities $S(Q)$ is
a monotonic function of $Q$ and has the shape of a `smoothed step
function' with values ranging from 
$\left(\pi^{3/2}\beta\varepsilon\rho\sigma^3\right)^{-1} \cong 0$
at low $Q$'s to unity at high $Q$'s. The crossover 
between the two regimes occurs
at a characteristic wavenumber $Q_{*}$ at which $S(Q) = 1/2$ and which,
according to eq.\ (\ref{sofq.analytic}), is given by
\begin{equation}
Q_{*}\sigma = 2
\sqrt{\ln\left(\pi^{3/2} \beta\varepsilon\rho\sigma^3\right)}.
\label{qstar}
\end{equation}
Note the very weak, 
square root-logarithmic dependence of $Q_{*}$ on density and the inverse
temperature.

Another 
quantity of interest is the isothermal compressibility $\chi_T$ of
the system, defined as
\begin{equation}
\chi_T = \left(V {{\partial^2F}\over{\partial V^2}}\right)^{-1},
\label{chi}
\end{equation}
where $F$ is the Helmholtz free energy,
and also related to the $Q \rightarrow 0$ limit of $S(Q)$ 
through \cite{hansen:mcdonald}
\begin{equation}
\rho k_BT \chi_T = S(0).
\label{sumrule}
\end{equation}
The expressions (\ref{ideal.excess})
and (\ref{dft.mfa}) yield the Helmholtz free energy in the high
density limit as
\begin{equation}
F = k_BTN\left[ \ln\left( \rho\Lambda^3\right) - 1\right]
    + {{1}\over{2}}N\pi^{3/2}\varepsilon\rho\sigma^3,
\label{fren}
\end{equation} 
and from eqs.\ (\ref{chi}) and (\ref{fren}) we obtain 
\begin{equation}
\chi_T = {{1}\over{k_B T \rho + \pi^{3/2}\varepsilon\rho^2\sigma^3}},
\label{chi.exact}
\end{equation}
which evidently satisfies the compressibility sum rule
(\ref{sumrule}) with $S(Q)$ given by eq.\ (\ref{sofq.analytic}). We note
that at high densities $\chi_T$ obeys the scaling
\begin{equation}
k_B T \chi_T \sim t \rho^{-2}.
\label{chi.scaling}
\end{equation}  
Moreover, from eq.\ (\ref{fren}) the pressure 
$P = -(\partial F/\partial V)$ is obtained as
\begin{equation}
P = k_BT\rho + {{1}\over{2}}\pi^{3/2}\varepsilon\rho^2\sigma^3.
\label{pressure}
\end{equation}

Since $S(Q) = 1 + \rho \tilde h(Q)$, eq.\ (\ref{sofq.analytic}) 
immediately yields an analytic expression for the Fourier transform
$\tilde h(Q)$ of the correlation function $h(r)$, namely
\begin{equation}
 \tilde h(Q) = -{{\pi^{3/2}\beta\varepsilon\sigma^3
                 \exp\left[-\left( Q\sigma/2 \right)^2\right]}
                 \over
                 {1 + \pi^{3/2}\beta\varepsilon\rho\sigma^3
                 \exp\left[-\left( Q\sigma/2 \right)^2\right]}}.
\label{hofq.analytic}
\end{equation}
At low $Q$'s, where the exponential is of order unity, the term proportional
to the density in the denominator dominates and $\tilde h(Q)$ behaves
as $-\rho^{-1} \rightarrow 0$ as $\rho \rightarrow \infty$. At high 
$Q$'s, the exponential itself tends to zero. Hence, the function
$\tilde h(Q)$ vanishes as $\rho \rightarrow \infty$ with the leading
term being proportional to $\rho^{-1}$. Consequently, the correlation
function $h(r)$ vanishes as well and $g(r) \rightarrow 1$ as
$\rho \rightarrow \infty$. This is the particular `high density 
ideal gas' limit of the model. However, a clear distinction must be
drawn between this 
`interacting ideal gas'
and the usual ideal gas, in which either the system is noninteracting
or an interacting system is considered at the opposite limit, 
$\rho \rightarrow 0$. In the usual ideal gas limit, we have
$c(r) = \exp[-\beta v(r)] - 1$, $S(Q) = 1$ and $g(r) = \exp[-\beta v(r)]$.
The ideal gas pressure $P$ scales linearly with the density and
the ideal compressibility $\chi_T$ scales with $\rho^{-1}$.
Here, $c(r) = -\beta v(r)$, $S(Q)$ is not unity in the whole 
$Q$ range, the pressure scales as $\rho^2$ [see eq.\ (\ref{pressure})]
and the isothermal compressibility as $\rho^{-2}$ 
[see eq.\ (\ref{chi.scaling})]. Nevertheless, the above considerations
point out to a kind of interesting {\it duality} of the GCM in the
liquid phase, in which the system is trivially ideal at low densities
and becomes again ideal (vanishing correlations) at high densities.
This can be thought of as
the counterpart for the fluid state of the duality 
discovered by Stillinger in the crystalline state 
\cite{stillinger:prb:79}.

The mean-field approximation put forward in this section is not limited
to the Gaussian potential. It should be valid for all interactions 
$v(r)$ which are finite, analytic functions and which tend to zero
fast enough at infinite separations so that the thermodynamic functions
(e.g., the internal energy)
are extensive and the Fourier transform
of $v(r)$ exists. Hence, we make the theoretical prediction that this
peculiar high-density ideal gas limit should exist for all potentials
satisfying the requirements put forward above. In addition, it can
now be understood
why the HNC showed the best agreement with simulation
results among all closures at high densities. As $g(r) \to 1$,
$h(r) \to 0$ and $c(r) \to -\beta v(r)$ in this limit, the exact
relation (\ref{gandc.exact}) forces the bridge function $B(r)$ to
vanish.
For bounded, analytic potentials
decaying fast enough to zero, the HNC becomes {\it exact} at high
densities.

\section{The solid: Einstein model\label{einstein}}

In this section we present the approach employed for the calculation
of the free energies of candidate crystalline states of the model,
necessary for determining the phase diagram of the GCM.
As we are dealing with a soft interaction, a harmonic approximation
in the solid is justified and we adopt the simple Einstein 
model \cite{ashcroft:mermin,tejero:daanoun:95}
as a means to estimate the free energy of the latter. The approach is
based on the Gibbs-Bogoliubov inequality \cite{hansen:mcdonald}. The
latter states that the 
Helmholtz free energy $F$ of a system having Hamiltonian
${\mathcal H}$ is related to the Helmholtz free energy $F_0$ of a
reference system having Hamiltonian ${\mathcal H}_0$ by 
\begin{equation}
F \leq F_0 + \langle {\mathcal H} - {\mathcal H}_0 \rangle_0,
\label{gibbs.bog}
\end{equation}
where the canonical
average on the right hand side is taken in the reference system.
The procedure is useful if (i) a simple enough reference Hamiltonian 
can be chosen, which physically corresponds to a situation 
close enough to the real one, and in which $F_0$ and the average
$\langle {\mathcal H} - {\mathcal H}_0 \rangle_0$ can be 
calculated in a straightforward way and (ii) this Hamiltonian contains
at least one variational parameter which can be chosen in order
to minimise the right hand side, obtaining in this way a lower upper
bound for the true free energy of the system.

In a harmonic solid, a reasonable non-interacting reference system is
the Einstein solid, characterised by the Hamiltonian
\begin{equation}
{\mathcal H}_0 = \sum_{i=1}^{N}
\left[{{{\bf p}_i^2}\over{2m}} + 
{{k}\over{2}} \left( {\bf r}_i - {\bf R}_i \right)^2\right],
\label{h0}
\end{equation}
where ${\bf p}_i$ is the canonical momentum of a particle of mass $m$,
$k$ is the `spring constant' which plays the role of a variational
parameter and the set $\{{\bf R}_i\}$ forms a prescribed Bravais lattice. 
For the GCM, the real Hamiltonian reads as
\begin{equation}
{\mathcal H} = \sum_{i=1}^{N} {{{\bf p}_i^2}\over{2m}} +
\sum_{i = 1}^{N}\sum_{j = i+1}^{N} \varepsilon
\exp\left(-{{|{\bf r}_i - {\bf r}_j|^2}\over{\sigma^2}}\right).
\label{h}
\end{equation} 

The calculation of the
Helmholtz free energy of the Einstein solid is a trivial exercise
yielding
\begin{equation}
{{F_0}\over{N}} = 
{{3}\over{2}}k_BT\ln\left({{\tilde\alpha}\over{\pi}}\right) +
3k_BT \ln\left({{\Lambda}\over{\sigma}}\right),
\label{f0}
\end{equation}
where 
\begin{equation}
\tilde\alpha = {{\beta k \sigma^2}\over{2}}.
\label{tildea}
\end{equation} 

The one- and two particle densities in the reference system are given
by a sum of Gaussians and a double sum of products of Gaussians,
respectively, where the latter are centered on lattice sites and
the sums are carried upon all these sites. As the interaction has itself
a Gaussian form, the calculation of
the quantity $\langle {\mathcal H} - {\mathcal H}_0 \rangle_0$
reduces to integrals that can be carried out analytically. The
final result reads as
\begin{equation}
{{\langle {\mathcal H} - {\mathcal H}_0 \rangle_0}\over{N}} = 
{{1}\over{2}}\sum_{j \ne 0} n_j
\left({{\tilde\alpha}\over{\tilde\alpha + 2}}\right)^{3/2}
\varepsilon\exp\left(-{{\tilde\alpha X_j^2}\over{\tilde\alpha + 2}}\right)
- {{3k_BT}\over{2}}.
\label{average}
\end{equation} 
The sum on the rhs is carried over all shells of lattice vectors,
i.e., sets of lattice vectors of equal length, with the shell 
$j = 0$ (self-interaction) excluded. The quantity $n_j$ is the
number of lattice vectors belonging to a shell and 
$X_j = |{\bf R}_j|/\sigma$. It can be easily seen that at $T = 0$,
where $\tilde\alpha \rightarrow \infty$, the first term on the rhs
of eq.\ (\ref{average}) reduces to the internal energy per particle
of the considered crystalline arrangement, the {\it lattice sum}.
At zero temperature there is no variational parameter and the 
Einstein model becomes exact; the winning phase is the one with
the lowest lattice energy. At finite temperatures, the sum of the
terms on the rhs of eqs.\ (\ref{f0}) and (\ref{average}) has to
be minimised with respect to $\tilde\alpha$ for any given lattice
structure. The minimum comes about through the competition of the
entropic, logarithmic term on the rhs of eq.\ (\ref{f0}), which
favors delocalisation at $\tilde\alpha = 0$ and the internal energy
term on the rhs of eq.\ (\ref{average}) which favors localisation
at $\tilde\alpha \rightarrow \infty$.
The obtained value is then the estimate for the Helmholtz
free energy of the given lattice and the procedure can be repeated
for every candidate lattice. We note that the term 
$3k_BT \ln(\Lambda/\sigma)$ on the rhs of eq.\ (\ref{f0})
can be dropped because it occurs for
all possible phases of the system, fluid and solid, and does not
affect the free energy comparisons between them.

We have performed the minimisation for a different number
of candidate lattice structures: fcc, bcc, simple cubic,
diamond and body-centered orthogonal, in which the
ratios between the lattice constants of the
conventional unit cells were used as additional variational
parameters \cite{watzlawek:etal:prl:99,martin:phd:00}. 
We always assumed a 
lattice with a single occupancy per site, based on the result
of Stillinger stating that indeed solids with multiply occupancies
are unstable \cite{stillinger:76}. For the whole range of
temperatures $0 \leq t \leq 0.015$ considered, we always found
the fcc and bcc to be the only stable crystals, with the former
winning at low densities and the latter at high densities. The
quantitative results are shown in section \ref{phdg}.
    
\section{Quantitative phase diagram\label{phdg}}

In this section we switch from the variable $\eta$ to the variable
$\bar\rho \equiv \rho\sigma^3$ as a measure of the density, in order
to allow for a direct comparison with the approximate results of
Stillinger and Stillinger \cite{stillinger:stillinger:97}.
With the free energy of the crystals obtained by the procedure 
outlined in section \ref{einstein}, the phase diagram can be drawn
if the corresponding free energy of the fluid is also known. The
latter can be obtained from the results of the integral equation
theories outlined in section \ref{iets}. 

The Helmholtz free energy of the liquid is the sum of the ideal and
excess terms (see section \ref{ideal}), namely
\begin{equation}
F(\rho,T) = Nk_BT \left[\ln\left(\rho\sigma^3\right) - 1\right] +
          3Nk_BT\ln\left({{\Lambda}\over{\sigma}}\right) +
          F_{\rm ex}(\rho,T).
\label{fliq}
\end{equation} 
If for a fixed temperature the radial distribution function $g(r)$ is
known for a region of densities, then one possibility to calculate the
free energy is through the so-called {\it virial route}. Here, one calculates
for every density the excess pressure $P_{\rm ex}$ of the system via
\begin{equation}
P_{\rm ex} = -{{2\pi\rho^2}\over{3}}
\int_0^{\infty} r^3 v'(r)g(r)dr,
\label{pex}
\end{equation}
where $v'(r) = dv(r)/dr$. Then, the excess free energy can be calculated 
by integrating the thermodynamic relation 
$P_{\rm ex} = -\partial F_{\rm ex}/\partial V$ from $\rho = 0$ up to 
the given density, under the initial condition $F_{\rm ex}(\rho = 0, T) = 0$.
Alternatively, one can use the results for the structure factor $S(Q)$
and the definition of the isothermal compressibility $\chi_T$
[eqs.\ (\ref{chi}) and ({\ref{sumrule})] to calculate the liquid free
energy, following in this way the {\it compressibility route}.  

If thermodynamic consistency between the two routes is not explicitly
enforced in an otherwise approximate closure, then the results from two 
routes are different, a problem known as {\it thermodynamic inconsistency}
of the closure \cite{hansen:mcdonald}.
The RY- and ZSEP-closures are thermodynamically 
consistent but the HNC is not. However, as explained in section \ref{iets},
neither the RY nor the ZSEP yield reliable results for the whole density
range. If we were dealing with a usual system, without reentrant melting,
then we could have used the RY or the ZSEP results at low or intermediate
densities. However, we are interested also in the high-density fluid
free energies, where in fact the HNC becomes exact. A combination of 
low-density results from one closure and high-density results from
another is not of much use either, because it would produce unphysical
discontinuities in the free energy or its derivatives at the point
of switching between the two. We are thus led to employ the HNC closure
in the whole density domain in order to perform the thermodynamic 
integration and obtain the fluid free energy.

The HNC compressibility route yields at low densities fluid free
energies that are too low, leading to the erroneous result that 
the HS-like freezing of the Gaussian fluid into a fcc lattice 
does not take place. There are two factors playing an important
role here: on the one hand, the predicted isothermal compressibilities
are too high causing a fluid free energy which is too low and on
the other hand, the solid free energy, being a product of the 
variational procedure outlined in section \ref{einstein}, is
unavoidably higher than the true one, see eq.\ (\ref{gibbs.bog}).
It is therefore pertinent
to follow the HNC virial route in calculating the fluid free energy.
The latter leads indeed to an overestimation of the fluid free energy
but this compensates for the overestimation of the solid free energy
and leads to the physically correct picture of freezing into an fcc
solid at the low-density part of the phase diagram. We have thus 
calculated the fluid free energies through the HNC virial route for a
range of temperatures $10^{-5} \leq t \leq 0.015$ for a range of
densities $0 \leq \rho\sigma^3 \leq 1.00$ and performed the common
tangent construction on the resulting 
$F_{\rm fluid}(\rho)/V$- and $F_{\rm solid}(\rho)/V$-curves to 
obtain the phase boundaries. The resulting phase diagram is 
shown in Fig.\ \ref{pd.plot}. 
\begin{figure}[hbt]
      \begin{center}
\includegraphics[width=12.0cm,height=10.0cm,clip]
{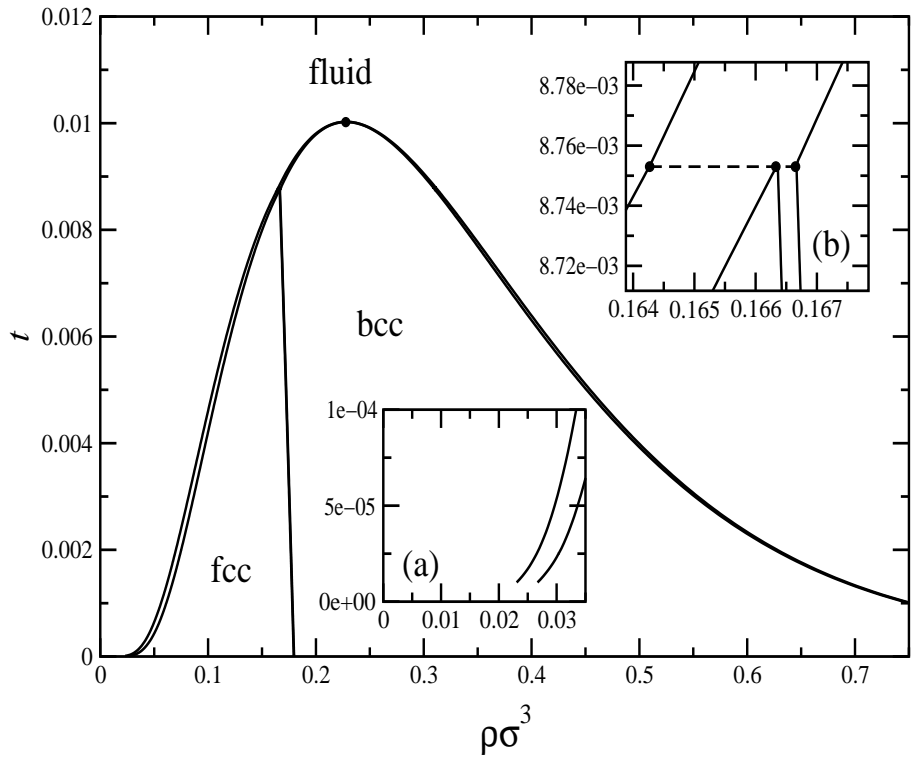}
      \end{center}
   \caption{The phase diagram of the GCM obtained by the approach described
            in the text. The fcc-bcc coexistence lines are also double
            lines but they cannot be resolved in the scale of the 
            figure because the fcc-bcc density gap is too small.
            The full dot marks the point at which the fluid-bcc 
            coexistence curves turn around. 
            The two insets show details of the phase diagram.
            (a) In the neighbourhood of zero densities and temperatures.
            (b) In the neighbourhood of the fluid-fcc-bcc triple temperature,
            with the dashed line denoting the triple line between these 
            coexisting phases.}
   \label{pd.plot}
\end{figure}

The phase diagram obtained is very similar to the approximate one
drawn by Stillinger and Stillinger \cite{stillinger:stillinger:97}.
It shows the sequence of freezing, structural (fcc $\to$ bcc) and
remelting transitions as well as the upper freezing temperature
$t_{\rm u}$ associated with the corresponding density 
$\bar\rho_{\rm u}$.
The coordinates of this point, where the fluid-bcc coexistence lines
turn around, are $(t_{\rm u}, \bar\rho_{\rm u}) = (0.0102, 0.2292)$. This
is in perfect agreement with the preliminary results from 
section \ref{iets}, where at $t = 0.01$ the structure factor
at $\eta = 0.12$, corresponding to $\bar\rho = 0.2292$, was found 
to slightly exceed the Hansen-Verlet value. The fcc-bcc coexistence
lines run linearly from the points $\bar\rho_{\rm fcc} = 0.17941$ and 
$\bar\rho_{\rm bcc} = 0.17977$ at $t = 0$ to the points
$\bar\rho_{\rm fcc} = 0.16631$ and $\bar\rho_{\rm bcc} = 0.16667$
at the triple temperature $t_{\rm t} = 8.753 \times 10^{-3}$.
The density gap between the fcc- and bcc-coexisting densities remains
constant and equal to $\Delta\bar\rho = 3.6 \times 10^{-4}$. The 
density of the coexisting fluid at the triple temperature is 
$\bar\rho_{\rm fluid} = 0.16431$. 

It should be emphasised that, notwithstanding
its deceiving appearance in Fig.\ \ref{pd.plot}, the
point $(t_{\rm u},\bar\rho_{\rm u})$
is {\it not} a critical point \cite{stillinger:76}.
At $(t_{\rm u},\bar\rho_{\rm u})$,
two common tangents between the fluid- and bcc-solid free energies,
one lying on the low- and the other on the high-density side of it,
coalesce into this single point. No susceptibility
diverges and all free energy density curves remain strictly concave up.

It is now pertinent to ask whether the Hansen-Verlet freezing criterion
is satisfied for both the low- and 
the high-density crystallisation of the
system. To this end, we have performed additional MC simulations at
temperatures below $t_{\rm u}$ and in   
Fig.\ \ref{ssim.lowt.plot} we show structure factors at 
two such temperatures, $t = 0.007$ and $t = 0.005$,
for increasing values of the density.
It can be seen that the Hansen-Verlet criterion is indeed valid
for both freezing transitions, a feature also observed for the
reentrant melting phenomenon in star polymer 
solutions \cite{watzlawek:etal:prl:99,martin:phd:00}.  
\begin{figure}[hbt]
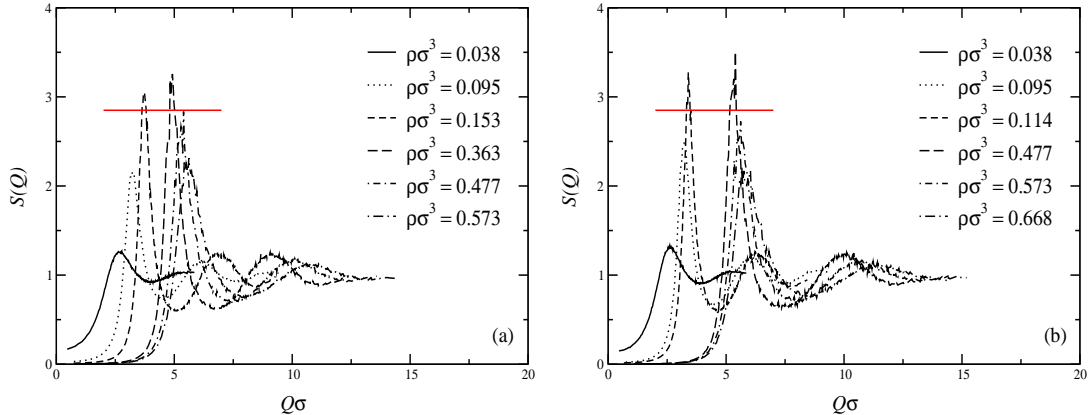

\begin{center}
\begin{minipage}[t]{7.2cm}
   \includegraphics[width=7.0cm,height=5.5cm,clip]
   {ssim.t0.007.eps}
   \end{minipage}
   \begin{minipage}[t]{7.2cm}
   \includegraphics[width=7.0cm,height=5.5cm,clip]
   {ssim.t0.005.eps}
   \end{minipage}
   \end{center}
\caption{Structure factors $S(Q)$ for the GCM at (a) $t = 0.007$ and
(b) $t = 0.005$ obtained from MC simulations. Densities as indicated
in the legends. The straight lines mark the Hansen-Verlet value 2.85. 
The corresponding structure factors in the regions 
$0.153 \lesssim \rho\sigma^3 \lesssim 0.363$ for $t = 0.007$
and $0.114 \lesssim \rho\sigma^3 \lesssim 0.477$ for $t = 0.005$
show Bragg peaks, indicating that the GCM has to be in the solid state
for these densities.}
   \label{ssim.lowt.plot}
\end{figure}

It is of course also possible to calculate the exact free energies of the 
various candidate phases in a simulation, by means, e.g., of the 
virial route in the fluid state and by employing the Frenkel-Ladd
method 
\cite{watzlawek:etal:prl:99,martin:phd:00,likos:pensph:98,frenkel:smit,frenkel:prl:86}
in the solid state. However, the latter is very time-consuming.
The very good agreement between the phase boundaries obtained from the 
approximate theory presented here and the MC results regarding
the height of the peak of $S(Q)$ and the spontaneous crystallisation
of the system in the simulation box, give us confidence that the
phase diagram of Fig.\ \ref{pd.plot} is quantitatively correct.

\section{Summary and concluding remarks\label{summary}}

The Gaussian core model displays a whole range of unusual phenomena and
properties: an anomalous dependence of the correlation functions on
the density, a high-density `ideal gas limit', reentrant melting and
an upper freezing temperature. Many of these characteristics arise
from the fact that the pairwise interaction does not diverge at zero
separation, i.e., it is {\it bounded}. However, the topology of
the phase diagram of the GCM has some striking similarities with
that of star polymers 
\cite{watzlawek:etal:prl:99,martin:phd:00}, obtained by
employing the {\it diverging} interaction given by 
eq.\ (\ref{interaction.star}). An upper freezing parameter 
($t$ in the GCM and $f$ in the star potential) appears in both,
above which the systems remain fluid at all densities. Freezing and
reentrant melting occur also at both systems. Yet, the phase diagram
of the GCM displays only two solid phases, whereas that of stars
has a richness of exotic crystal structures \cite{watzlawek:etal:prl:99}.
The latter are caused by the crossover of the star potential from a
logarithmic to a Yukawa form, a feature absent in the GCM.

To the best of our knowledge, the only other bounded interaction
that has been studied by similar techniques to-date is the 
penetrable spheres model (PSM) mentioned in section
\ref{gaussian} \cite{likos:pensph:98,fernaud:jcp:00,schmidt:cecam:99}.
The phenomenology and the associated phase diagram of the PSM are
quite different from those of the GCM. These two bounded potentials
have phase diagrams which do not look at all similar to one another.
This comes as a striking difference to the relative insensitivity
of the phase diagrams of unbounded interactions.
In the PSM 
no reentrant melting occurs and the system seems to
freeze at {\it all} temperatures into increasingly
clustered solids \cite{fernaud:jcp:00,schmidt:cecam:99}. 
The reason for this difference is, evidently, that in the PSM, the
particles can build clusters.
In the GCM, this
mechanism is not present; the interaction varies rapidly enough with
distance, so that multiple occupancies are penalised. 
Therefore, the boundedness of the interaction brings about a 
new factor to be considered:
multiple occupancies, which are prohibited from the very beginning
for diverging interactions, have to always been taken into
account whenever one deals with bounded ones.

An important conclusion which appears to be valid for a large class
of bounded potentials has been nevertheless drawn, and it is the
mean field-ideal gas behaviour of all such systems at high densities.
The PSM does not belong to this class; indeed, the discontinuity 
of the PSM pair potential at $r = \sigma$
forces its radial distribution function
to have a jump at the same position
at all densities and hence an ideal gas limit can never
be attained in this model. This is a feature complementary but not
identical to the clustering property of the system. On the other hand,
one can easily construct bounded pair interactions depending on
some parameter, so that the sharpness of the decay of the PSM from
a finite to zero value can be tuned. It will be very interesting
to examine the structure and thermodynamics
of such a family of systems as a function
of the `smoothening parameter' and establish the limits of the
clustering- and the ideal gas-behaviour at high densities. We plan
to return to this problem in the future.

\ack

We thank Joachim Dzubiella, Mar{\'\i}a Jos{\'e} Fernaud, Prof J-P Hansen
and Dr Ard Louis for
helpful discussions, 
and Arben Jusufi for providing Fig.\ \ref{twochains.plot}.
This work was supported by the \"Osterreichische Forschungsfond under Project
Nos P11194-PHY and P13062-TPH. 
AL acknowledges financial support 
by the Deutsche Forschungsgemeinschaft within the 
Sonderforschungsbereich (SFB) 237.

\end{document}